\documentclass[a4paper,11pt]{article}
\usepackage[a4paper, total={17cm, 25cm}]{geometry}
\usepackage{fancyhdr}
\fancypagestyle{plain}{
\fancyhf{} 
\fancyhead[RH]{\thepage}
\fancyfoot[CF]{28th Texas Symposium on Relativistic Astrophysics \\ Geneva, Switzerland -- December 13-18, 2015}

}
\usepackage[english]{babel}
\usepackage{amssymb,amsmath}
\usepackage{endnotes}
\usepackage{graphicx}
\begin{document}

\title{\bf The Blandford-Znajek Theory Revisited}

\author{
Luigi Foschini, INAF -- Osservatorio Astronomico di Brera, Merate, Italy
}

\date{January 15, 2016}

\maketitle


\section{Introduction}
It is today commonly known that accreting black holes (BHs) eject matter at relativistic speeds in the form of bipolar jets. The jet itself is believed to be an important relief valve for the angular momentum, thus enhancing the accretion of matter onto the central object. The wide variety of black hole masses ($\sim 3-10^{10}M_{\odot}$) and environments where jets develop (from Galactic binaries to Active Galactic Nuclei, AGN), suggests the presence of a universal engine able to work under the most disparate conditions. As noted by Mario Livio, while the physics of the jet formation is likely to be always the same, the appearance depends on the source \cite{LIVIO}. Indeed, as proved in \cite{FOSCHINI}, powerful relativistic jets in Galactic binaries and AGN can be unified according to the scaling theory by \cite{HEINZ}.

Nevertheless, how relativistic jets are generated remains a hot topic of the present-day astrophysics. As Roger Penrose outlined the possibility to extract energy from a black hole \cite{PENROSE}, many other authors proposed their theories. One of the most important results was published by Blandford \& Znajek (BZ, \cite{BZ, MDT}), after some interesting ``precursors'' such as Ruffini \& Wilson \cite{RUFFINI} and Lovelace \cite{LOVELACE}. A long-lasting debate about the physics of the BZ theory started in nineties, as Punsly \& Coroniti \cite{PC} casted doubts about the causality of the magnetic field threading the event horizon. Although the basic concepts are now almost clarified (\cite{BESKIN, KOM1, KOM2}), the BZ theory still anchors the creativity of theorists (e.g. \cite{GRALLA, KOIDE, TOMA}). 

Aim of the present work is to add one more point of view. There are indeed some interesting and intriguing features of the BZ theory that are relatively unexplored. 

\section{The power of the slip}
The power extracted from a rotating black hole via the BZ process can be summarized by the following equation:

\begin{equation}
L_{\rm BZ} = k f(B,r_{\rm H},a) \frac{\omega_{\rm F}(\omega_{\rm H}-\omega_{\rm F})}{\omega_{\rm H}^2}
\label{EQ:BZ}
\end{equation}

\noindent where $k$ is a normalization constant; $f$ is a function of the event horizon radius $r_{\rm H}$, the intensity of the magnetic field $B$, the spin $a$; $\omega_{\rm H}$ is the angular speed of the black hole, while $\omega_{\rm F}$ is that of the magnetic field lines. The latter depends on the type and geometry of the accretion disk \cite{GHOSH, MODERSKI}. 

In Eq.~(\ref{EQ:BZ}), I have emphasised the term composed by the angular speeds, which is the core of the BZ engine as stressed by \cite{KOM1, KOM2}. Since most of the authors focus on the jet generation, this term was often studied for this aim only (e.g. \cite{BESKIN, MARASCHI, AT}), set to its optimal value $\omega_{\rm F}=0.5\omega_{\rm H}$ \cite{MDT}, and the other values neglected. A few authors gave a quick glimpse on what changes when $\omega_{\rm F}\lessgtr\omega_{\rm H}$ \cite{LI, LIVIO2, UZDENSKY1, UZDENSKY2, WANG}, while others explored the analogy with an electric machine (\cite{LOVELACE}, \cite{LI}, \cite{PHINNEY}, \cite{WANG}, but see \cite{KOM2}). 

Here I would like to expand the study on the effects of the relative angular velocity between the black hole and the magnetic field lines. Let us to define the {\it slip factor} $s$ (or simply {\it slip}) as:

\begin{equation}
s = \frac{\omega_{\rm F}-\omega_{\rm H}}{\omega_{\rm F}} = 1 - \frac{\omega_{\rm H}}{\omega_{\rm F}}
\label{EQ:SLIP}
\end{equation}

\noindent which implies that -- in the view of the electric machine analogy -- I have considered the black hole as the ``rotor'' and the magnetic field lines as the ``stator''. Obviously the latter does not need to be really at rest: what is important is the relative motion between the two structures. Moreover, these terms are just for consistency with the discussion below, when I will speak about generator and motor. It is well-known that the event horizon has no magnetic field, but -- as noted by \cite{KOM1, WALD} -- there is a rotationally-induced electric field. 

Then, the Eq.~(\ref{EQ:BZ}) can be rewritten as a function of the slip:

\begin{equation}
L_{\rm BZ} \propto \frac{s(s-1)}{(1-s)^3}
\label{EQ:BZSLIP}
\end{equation}

\noindent and in the following I refer to the right-hand side of the equation as the {\it slip term}. I underline that I am not interested here in the estimate of the intensity of the BZ power, whose value is often debated (see e.g. \cite{GHOSH, LIVIO2, AT}), but rather at the direction of its flow, which in turn is driven by the sign of the {\it slip term} of the Eq.~(\ref{EQ:BZSLIP}). Therefore, I do not deal with either the normalization constant $k$ or the function $f$, but I focus on the relationship of $L_{\rm BZ}$ with the slip $s$. 

\begin{figure}[t]
\begin{center}
\includegraphics[angle=270,scale=0.5]{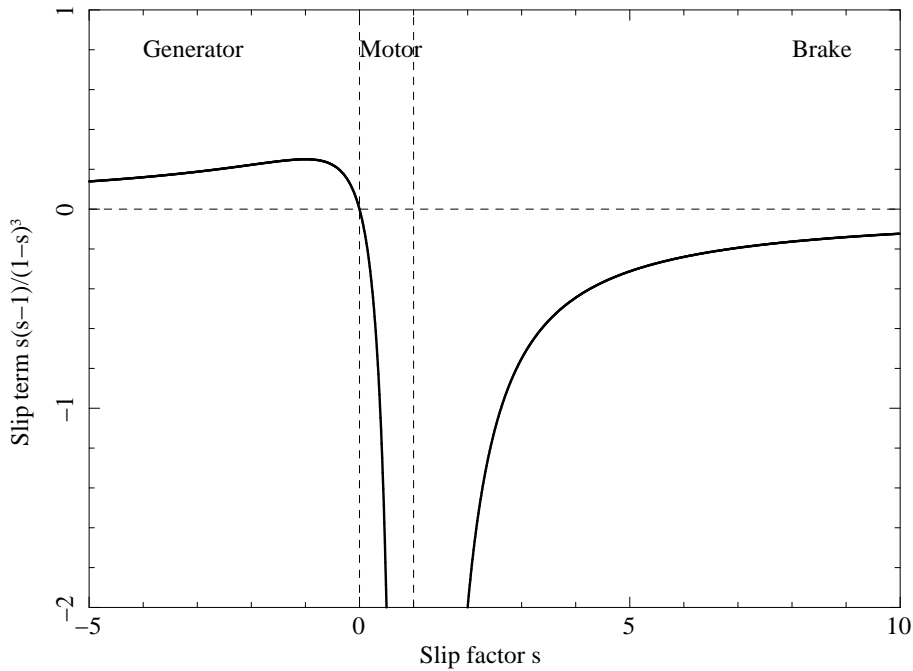}
\caption{Slip term of the BZ power in the Eq.~(\ref{EQ:BZSLIP}) as a function of the slip factor. See the text for details.}
\label{FIG:SLIP}
\end{center}
\end{figure}

The Eq.~(\ref{EQ:BZSLIP}) is displayed in Fig.~\ref{FIG:SLIP} as a function of $s$ and it is possible to distinguish some interesting cases:

\begin{itemize}
\item[(i)] $\omega_{\rm H}/\omega_{\rm F}=1 \Rightarrow s=0$: it is the {\it synchrony} condition, i.e. the angular speed of the black hole and that of the magnetic field lines are the same. This results in null BZ power ($L_{\rm BZ}=0$).

\item[(ii)] $\omega_{\rm H}/\omega_{\rm F}=0 \Rightarrow s=1$: the black hole is at rest (Schwarzschild case) and $L_{\rm BZ}\rightarrow -\infty$, like a short circuit where the power is completely drained by the black hole. 

\item[(iii)] $\omega_{\rm H}/\omega_{\rm F}=2 \Rightarrow s=-1$: this is the well-known condition of maximum output of the jet power (\cite{BZ,MDT}). The slip term in Eq.~(\ref{EQ:BZSLIP}) is equal to $0.25$.

\item[(iv)] $\omega_{\rm H}>\omega_{\rm F} \Rightarrow s<0$: the black hole angular speed is greater than that of magnetic field lines. The machine works as {\it generator} and the BZ power is positive ($\rightarrow$ jet emission). The rotational power of the black hole is converted into the electromagnetic jet power. 

\item[(v)] $\omega_{\rm H}<\omega_{\rm F} \Rightarrow 0<s<1$: the BZ power is negative, the machine works as a {\it motor}. The rotational energy of the magnetic field, which rotates faster than the BH, is extracted to power the motion of the black hole. If the BH increases its angular speed sufficiently to surpass that of the field, then it changes to the condition (iv).

\item [(vi)] $\omega_{\rm H}/\omega_{\rm F}<0 \Rightarrow s>1$: the magnetic field and the hole are counterrotating. The BZ power is negative and the machine works as a {\it brake}. As $s\rightarrow +\infty$, $L_{\rm BZ}\rightarrow 0$.
\end{itemize}

I would like to stress again that all the above considerations are valid {\it independently} on the intensity of the BZ power. The {\it slip term} due to $\omega_{\rm H}$ and $\omega_{\rm F}$ drives the direction of the BZ power. With respect to previous studies, the part for $\omega_{\rm H}<\omega_{\rm F}$ (negative BZ power) is now divided into two parts, the motor and the brake.

\section{Applications}
Past studies focused only on the formation of the relativistic jet and discarded all the other solutions. However, although the BZ theory has been developed to explain the relativistic jets, it can be broadly understood as the coupling of a black hole with its magnetosphere and accretion disk as outlined by the cases above. 

It is now possible to try explaining some measurements. The present day observations are often contradictory (e.g. \cite{SPIN1,SPIN2,FENDER}) and, since it is not (yet?) known how to measure $\omega_{\rm F}$, there is the temptation to directly link the BH spin with the presence of the jet. However, the observation of a maximally rotating BH in an AGN without jet \cite{WILMS} raised a severe challenge to this conjecture. 

The present study offers the possibility to explain the presence of a maximally rotating BH in an AGN without jet: it could be either the case (i) or (v). In the case reported by \cite{WILMS}, where the radius of the innermost stable orbit indicates that BH and disk (and hence the magnetosphere) are co-rotating, implies that (vi) cannot be applied. 

Thus, the dichotomy of AGN with or without jets can be explained as follows: AGN with no jets have the disk rotating faster than the BH or counterrotating, while the jet is present only when the BH rotates faster than the disk. In the case of Galactic binaries, the emission of the jet is related to the accretion state (low/hard, high/soft), which in turn can be explained in the same way: perhaps, disks in high state rotate faster than the BH, while the opposite occurs in low states. 

The main message that I would like to spread around is to stop neglecting the different cases of the BZ theory just because they are not producing jets. The BZ theory can have applications wider than expected. A broader study of the feedback between the black hole, the magnetosphere, and the accretion disk can offer interesting surprises in understanding these intriguing cosmic systems.



\begin{thebibliography}{99}
\bibitem{BESKIN} Beskin, V.~S. \& Kuznetsova, I.~V. (2000) Nuovo Cimento B 115, 795.

\bibitem{BZ} Blandford, R.~D. \& Znajek, R.~L. (1977) MNRAS 179, 433.

\bibitem{FENDER} Fender, R.~P. et al. (2010) MNRAS 406, 1425.

\bibitem{FOSCHINI} Foschini, L. (2014) Int. J. Mod. Phys. Conf. Series 28, 1460188.

\bibitem{GHOSH} Ghosh, P. \& Abramowicz, M.~A. (1997) MNRAS 292, 887.

\bibitem{GRALLA} Gralla, S.~E. \& Jacobson, T. (2014) MNRAS 445, 2500.

\bibitem{HEINZ} Heinz, S. \& Sunyaev, R.~A. (2003) MNRAS 343, L59

\bibitem{KOIDE} Koide, S. \& Baba, T. (2014) ApJ 792, 88.

\bibitem{KOM1} Komissarov, S. (2004) MNRAS 350, 427.

\bibitem{KOM2} Komissarov, S. (2009) J. Korean Phys. Soc. 54, 2503.

\bibitem{LI} Li, L.-X. (2000) ApJ 533, L115.

\bibitem{LIVIO} Livio, M. (1997) ``The formation of astrophysical jets''. In: Accretion phenomena and related outflows - IAU Colloqium 163, eds D.~T. Wickramasinghe, L. Ferrario, \& G. V. Bicknell, ASP Conference Series 121, p. 845.

\bibitem{LIVIO2} Livio, M., et al. (1999) ApJ 512, 100.

\bibitem{LOVELACE} Lovelace, R.~V.~E. (1976) Nature 262, 649.

\bibitem{MDT} Macdonald, D. \& Thorne, K.~S. (1982) MNRAS 198, 345.

\bibitem{MARASCHI} Maraschi, L., et al. (2012) J. Phys. Conf. Series 355, 012016.

\bibitem{SPIN1} McClintock, J.~E., et al. (2011) Class. Quantum Grav. 28, 114009.

\bibitem{MODERSKI} Moderski, R. \& Sikora, M. (1996) MNRAS 283, 854.

\bibitem{SPIN2} Narayan, R. \& McClintock, J.~E. (2012) MNRAS 419, L69.

\bibitem{PENROSE} Penrose, R. (1969) Riv. Nuovo Cimento 1, 252.

\bibitem{PHINNEY} Phinney, E.~S. (1983) ``Black hole - driven hydromagnetic flows''. In: Astrophysical Jets - Proceedings of the International Workshop, eds A. Ferrari \& A.~G. Pacholczyck, NATO-ASI Series 103, D. Reidel Publishing Co., Dordrecht, p. 201.

\bibitem{PC} Punsly, B. \& Coroniti, F.~V. (1990) ApJ 354, 583.

\bibitem{RUFFINI} Ruffini, R. \& Wilson, J.~R. (1975) Phys. Rev. D 12, 2959.

\bibitem{AT} Tchekhovskoy, A., et al. (2010) ApJ 711, 50.

\bibitem{TOMA} Toma, K. \& Takahara, F. (2014) MNRAS 442, 2855.

\bibitem{UZDENSKY1} Uzdenski, D. (2004) ApJ 603, 652.

\bibitem{UZDENSKY2} Uzdenski, D. (2005) ApJ 620, 889.

\bibitem{WALD} Wald, R.~M. (1974) Phys. Rev. D 10, 1680.

\bibitem{WANG} Wang, D.~X., et al. (2002) MNRAS 335, 655.

\bibitem{WILMS} Wilms, J., et al. (2001) MNRAS 328, L27.
\end{thebibliography}
\end{document}